\documentclass[aps,amsmath,amssymb,superscriptaddress,showpacs,twocolumn,floats,pdff]{revtex4}

\usepackage{amssymb,amsmath}
\usepackage{epsfig}

\newlength{\upit}\upit=0.1truein

\newcommand{\ltappr}{{{\lower4pt\hbox{$<$} } \atop \widetilde{ \ \ \ }}}
\newlength{\bxwidth}\bxwidth=1.5 truein

\newcommand{\beg}{\begin{equation}}
\newcommand{\en}{\end{equation}}

\newcommand{\bra}[1]{\langle {#1} |}
\newcommand{\ket}[1]{| {#1} \rangle}

\DeclareMathAlphabet{\mathpzc}{OT1}{pzc}{m}{it} 

\newlength{\figwidth}
\newlength{\shift}
\usepackage{color,ulem}

\newcommand \bea {\begin{eqnarray} }
\newcommand \eea {\end{eqnarray}}
\newcommand{\bk}{{\bf{k}}}

\begin{document}

\title{Surface Theory of a Family of Topological Kondo Insulators}

\author{Bitan Roy}
\affiliation{Condensed Matter Theory Center, Department of Physics, University of Maryland, College Park, MD 20742, USA}

\author{Jay D. Sau}
\affiliation{Condensed Matter Theory Center, Department of Physics, University of Maryland, College Park, MD 20742, USA}
\affiliation{Joint Quantum Institute and Condensed Matter Theory Center, Department of Physics, University of Maryland, College Park, Maryland 20742, USA}

\author{Maxim Dzero}
\affiliation{Department of Physics, Kent State University, Kent, OH 44242, USA}
\affiliation{CFIF, Instituto Superior T\'{e}cnico, Universidade de Lisboa, Av. Rovisco Pais, 1049-001 Lisboa, Portugal}

\author{Victor Galitski}
\affiliation{Condensed Matter Theory Center, Department of Physics, University of Maryland, College Park, MD 20742, USA}
\affiliation{Joint Quantum Institute and Condensed Matter Theory Center, Department of Physics, University of Maryland, College Park, Maryland 20742, USA}

\begin{abstract}
A low-energy theory for the  helical metallic states, residing on the surface of cubic topological Kondo insulators, is derived. Despite our analysis being primarily focused on a prototype topological Kondo insulator, Samarium hexaboride (SmB$_6$), the surface theory derived here can also capture key properties of other  heavy fermion topological compounds with a similar underlying crystal structure. Starting from an effective mean-field eight-band model in the bulk, we  arrive at a low-energy description of the surface states, pursuing both analytical and numerical approaches. In particular, we show that helical Dirac excitations occur near the $\bar{\Gamma}$ point and the two $\bar{X}$-points of the surface Brillouin zone and generally the energies of the Dirac points display {\it offset} relative to each other. We calculate the dependence of several observables (such as bulk insulating gap, energies of the surface Dirac fermions, their relative position to the bulk gap, etc.) on various  parameters in the theory.  We also investigate the effect of  a spatial modulation of the chemical potential on the surface spectrum and show that this band bending generally results in ``dragging down'' of the Dirac points deep into the valence band and strong enhancement of Fermi velocity of surface electrons. Comparisons with recent ARPES and quantum oscillation experiments are drawn.
\end{abstract}

\date{\today}


\maketitle

\section{Introduction}

Samarium hexaboride (SmB$_6$) has recently emerged as a prominent candidate for an ideal time-reversal- and inversion-invariant topological insulator -- a material which is insulating in the bulk, but hosts topologically protected metallic surface \cite{Theory1, Theory2, exp1, exp2, exp3, exp4, exp5, xia-fisk-NatMat, RecentOneDim}. The hallmark signatures of these gapless surface states are the {\it helical spin structure} and their robustness against time-reversal invariant perturbations \cite{Hasan,Moore}. In SmB$_6$, the {\it hybridization} between the conduction electrons occupying $d$ orbitals and predominantly localized electrons residing on $f$ orbitals drives an insulating gap opening at low temperatures. What also makes SmB$_6$ special is the presence of strong on-site Hubbard interaction between the samarium $f$-electrons \cite{Geballe,Allen,Cooley}. In particular, the Hubbard interaction is strong enough to favor the valence configuration with odd number of electrons, $4f^5$ and the hybridization between the conduction and  $f$ electrons drives the system into a mixed-valence regime between $4f^5$ and $4f^6$ configurations \cite{Peter}. 
\begin{figure}[htb]
\includegraphics[width=4.25cm,height=4.5cm]{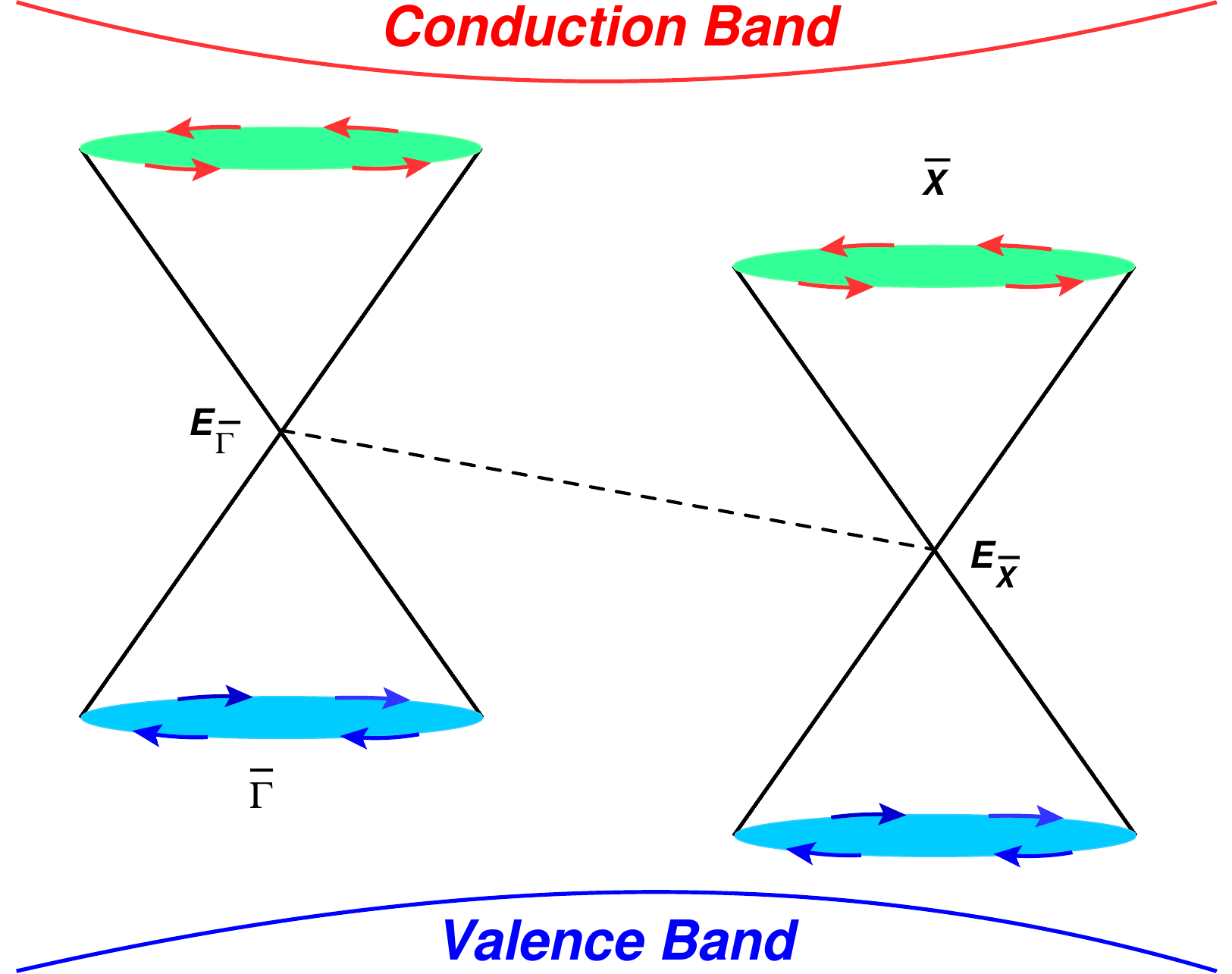}
\includegraphics[width=4.25cm,height=4.5cm]{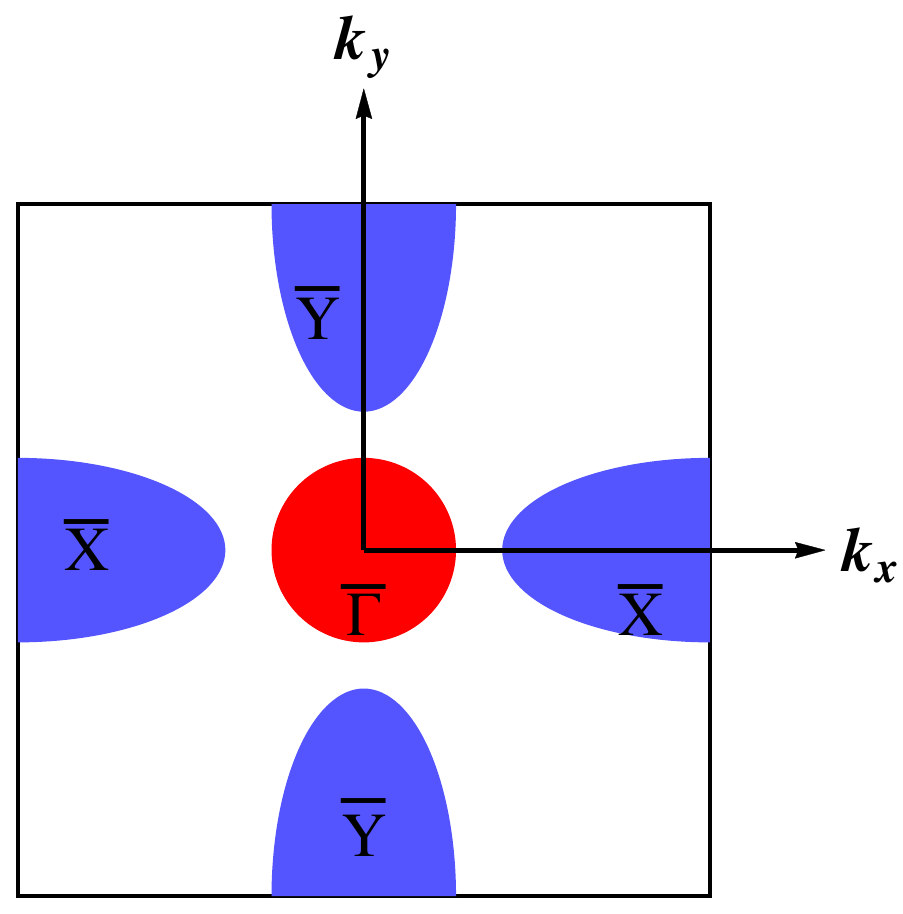}
\caption{(Color online) Left: position of the Dirac points at the $\bar{\Gamma}$ and $\bar{X}$ relative to the valence and conduction bands. It is in principle possible to drag down one or both the Dirac points into the valence band, by tuning some parameters in the effective theory, defined in Eq. (\ref{Hbulk}) or introducing a band bending potential, shown in Eq. (\ref{BBPOt}). $E_{\bar{\Gamma}}$ and $E_{\bar{X}}$ are the energies of the surface Dirac fermions at $\bar{\Gamma}$ and $\bar{X}$ of the surface BZ (see right panel). The dotted line represents a generic offset among these two Dirac points. The arrows represent the helical texture of the spin of the surface states. For the sake of simplicity we here assume that all three surface Dirac cones have identical chirality, and the corresponding spin-texture constitutes {\it vortex} in the momentum space. Right: a representative surface BZ, when the chemical potential ($\mu_s$) at the surface resides in between two Dirac points at $\bar{\Gamma}$ and $\bar{X}$ points, i.e., $ E_{\bar{X}}<\mu_s<E_{\bar{\Gamma}}$. The red and blue pockets, respectively, correspond to hole- and electron-like Fermi pockets. However, nature of the Fermi pockets depends on the location of $\mu_s$ with respect to $E_{\bar{\Gamma}}$ and $E_{\bar{X}}$. }\label{Fig1}
\end{figure}

While this theoretical work mainly concentrates on SmB$_6$,  we allude to its possible generalization  to address electronic properties of other cubic Kondo insulators (including other possible topological insulators in hexaboride family). Due to the presence of the strong electronic correlations in SmB$_6$ all the recent analytical approaches of computing the $Z_2$ topological invariant are based on either effective low-energy approximations \cite{Theory1,Theory2,Takimoto} or various types of large-$N$ mean-field theories \cite{KimN,LargeN,CTKI,Sigrist}. Generally, the main outcome of these studies is that a topologically nontrivial insulating state emerges due to the odd number of $d$- and $f$-bands inversions at the high-symmetry points of the Brillouin zone (BZ). First-principle calculations \cite{XiDai} as well as studies based on dynamical mean-field theory (DMFT) \cite{DMFT} also confirm this result \cite{xaidai-DFT}. In addition, the existence of various topologically distinct phases has been predicted from the DMFT analysis, which, for example, can be accessed by continuously tuning the strength of the on-site Hubbard interaction of the $f$-electrons from $U_{ff}=0$ to $U_{ff} \to \infty$. Since the transition between the two topologically distinct states must necessarily be separated by a gapless phase, trivial band insulators and topological Kondo insulators cannot be connected adiabatically \cite{MartinAllen1, Martin}.

The appearance of topologically nontrivial states in the $f$-electron insulators stems from the fact that the hybridization between the $d$- and the $f$-electrons necessarily needs to be an {\it odd function of momentum} to preserve the time-reversal and inversion symmetries. Therefore, the hybridization matrix element vanishes at the {\it high-symmetry} points of the BZ. Consequently, the $Z_2$ topological invariant is determined by the relative position of the renormalized $f$-electron (due to the Hubbard interaction) and conduction $d$-electron energies computed at the high-symmetry points of the BZ (see the Appendix). In particular, for a wide range of the parameters, corresponding to an average valence on Samarium, even and odd parity bands invert at the three X points of the BZ, suggesting that a three-dimensional topological insulating state can be realized in SmB$_6$. Note that band inversion at the X points implies the existence of  three Dirac points on the surface; one at $\overline{\Gamma}$ (in red) and two at $\overline{X}$ and $\overline{Y}$ (in blue) points of the two-dimensional surface BZ, as shown in Fig.~\ref{Fig1}, which has  been confirmed experimentally through a number of ARPES measurements \cite{exp4,exp5,ARPES2,ARPES3,ARPES4,ARPES5,ARPES6}. Interestingly, a similar surface band structure has been observed in another hexaboride compound - YbB$_6$ \cite{yb6ARPES, zahedybb6, ybb6-swissARPES}, although the underlying interaction-induced mechanism of the possible topological behavior has been argued to be different from Kondo hybridization \cite{zahedybb6, ybb6}.

In this paper we derive an effective model for the surface states in prototype cubic topological Kondo insulators, on the surfaces perpendicular to the main axes. Our effective surface model is derived from the bulk Hamiltonian, which takes into account a realistic band structure of SmB$_6$ \cite{CTKI}. Otherwise, near all three Dirac points, namely at the $\overline {\Gamma}$, $\overline{X}$, and $\overline{Y}$ points of the surface BZ, the effective low-energy description of the surfaces is captured by two-dimensional massless Dirac Hamiltonians. In the vicinity of the $\overline{\Gamma}=(0,0)$ point it goes as 
\begin{equation}\label{hsurfintroG}
H^{\bar{\Gamma}}_{sur}= v^{\bar{\Gamma}}_F \left( \sigma_x k_x -\sigma_y k_y\right),
\end{equation}  
representing an {\it isotropic} conical dispersion, where $\vec{k}$ is measured from the $\overline{\Gamma}$ point. On the other hand, in the vicinity of the $\overline{X}=(\pi,0)$ and $\overline{Y}=(0,\pi)$ points, the two-dimensional Dirac Hamiltonian is 
\begin{equation}\label{hsurfintroX}
H^{j}_{sur}= \left( v^{j}_{x} \sigma_x k_x - v^{j}_{y} \sigma_y k_y\right).
\end{equation}  
for $j=\overline{X}, \overline{Y}$ and generically $v^j_{x} \neq v^j_{y}$. In addition, we show that $v^{\overline{X}}_{x}= v^{\overline{Y}}_{y}$, and $v^{\overline{X}}_{y}= v^{\overline{Y}}_{x}$, reflecting the underlying {\it cubic symmetry} in the bulk of the system. Therefore, in the vicinity of $\overline{X}$ and $\overline{Y}$ points the conical Dirac dispersions are {\it anisotropic}. A representative two-dimensional surface BZ and the {\it helical spin texture} of low-energy quasiparticles are shown in Fig.~\ref{Fig1}. Although the spin textures near $\bar{\Gamma}$, $\bar{X}$, and $\bar{Y}$ in Fig.~1 corresponds to vortices in the surface BZ, the ones associated with $H_{\bar{\Gamma}}$, $H_{\bar{X}, \bar{Y}}$ in Eqs.~(\ref{hsurfintroG}) and (\ref{hsurfintroX}) respectively corresponds to {\it anti-vortices} in the momentum space. Nevertheless, both situations are protected by bulk strong $Z_2$ topological invariant. These features are in qualitative agreement with a number of ARPES measurements in SmB$_6$, and we obtain such low energy description of the surface state both analytically as well as numerically (see Sec. III). A subsequent mean-field theory approximation for the bulk Hamiltonian is controlled by the parameter $1/N$ with $N=4$ for SmB$_6$ corresponding to the four-fold degenerate $f$-orbital multiplet \cite{CTKI}. We here also determine the effective Fermi velocities, location of the Dirac points ($E_{\bar{\Gamma}}$ and $E_{\bar{X}}$ in Fig.~1), and penetration depth of the surface states for each of the three Dirac cones. When possible, we obtain closed analytical expressions for these quantities as a function of various microscopic parameters, appearing in the effective theory, describing a bulk Kondo-insulating state. In particular, we find that the values of the Fermi velocities are primarily controlled by the {\it renormalized} strength of the hybridization amplitude (due to the particle-hole anisotropy in the bulk) between $d$- and $f$-states on the surface.

Our method of finding the effective theory on the surface is similar to the one used to derive the model for surface states in Bi-based topological insulators \cite{Zhang1, Zhang2, Surface1, fanzhang-ST} - systems where electronic correlations are weak. Our main assumption in the first part of the paper is that the self-consistent mean-field theory for the `bulk plus surface' system will not significantly modify the values of the hybridization and chemical potentials compared to the mean-field theory for the bulk system only. In other words, we  assume there that the boundary does not significantly affect the parameters of the bulk. However, the non-universal boundary effects resulting in {\it band bending} are also considered later (see Sec.~\ref{bending}), by introducing a spatially modulated profile of the chemical potential for the $f$ electrons, and it is demonstrated that the band bending can qualitatively modify the surface band structure \cite{kittel, BBBiSe-1, BBBiSe-2, coleman-private, coleman-APS, 1d-bandbending}. We here show that in the presence of spatially modulated chemical potential, the Dirac points at $\bar{\Gamma}$ and $\bar{X}$ points can be gradually dragged down into the valence band, when its characteristic decay length into the bulk ($\lambda_{B}$) and/or its magnitude ($U_0$) is large enough. In addition, we find that the Dirac point at the $\bar{\Gamma}$ point gets immersed into the valence band for relatively weaker modulation of the chemical potential, while that near the $\bar{X}$ point continues to live inside the bulk Kondo insulating gap for a wider range of $\lambda_B$ and $U_0$ (see Figs. \ref{bandbend1}, and \ref{bandbend2}). Such peculiar behavior arises from the fact that the penetration depth for the surface state near the $\bar{\Gamma}$ point is smaller than that near the $\bar{X}$ point.

This paper is organized as follows. In the next section we formulate the effective tight-binding model for cubic topological Kondo insulators, which may serve as minimal model in various hexaboride compounds at low energies, and discuss the bulk band structure. In Sec. III, we explicitly derive the surface states and obtain surface Hamiltonians. In this section, we also present the band structure of the surface BZ, and demonstrate the explicit dependence of various quantities such as Fermi velocity, energies of the Dirac fermions, penetration depths etc., of the surface states on the band parameters. Section IV is devoted to address the effect of spatial modulation of the chemical potential or the band bending on the structure of the surface states. In Sec. V we summarize our main findings and compare the results with recent ARPES and quantum oscillation measurements. We show the computation of the bulk topological invariant within the framework of our effective minimal model in the Appendix.


\section{Model Hamiltonian in bulk}
\label{bulk}

Let us first introduce the effective tight-binding or mean-field model for the cubic Kondo insulators, with our focus being on a prototype system, SmB$_6$ \cite{CTKI}. SmB$_6$ has a simple cubic structure with a clusters of six boron (B) atoms located at the center of the unit cell, acting as spacers which mediate electronic hopping among the samarium (Sm) sites. Recent ``LDA + Hubbard-U" band structure calculations suggest that the Kondo hybridization is {\it strongest} between samarium 4f orbitals and dispersing $d$ bands which form electron pockets around the X points of the BZ \cite{Antonov}. Based on these predictions, we wish to promote here an effective model for a family of topological Kondo insulators, which share similar underlying cubic symmetry of SmB$_6$, such as PuB$_6$, for example \cite{kottlier}.  

\begin{figure}[htb]
\includegraphics[width=8.25cm,height=6.5cm]{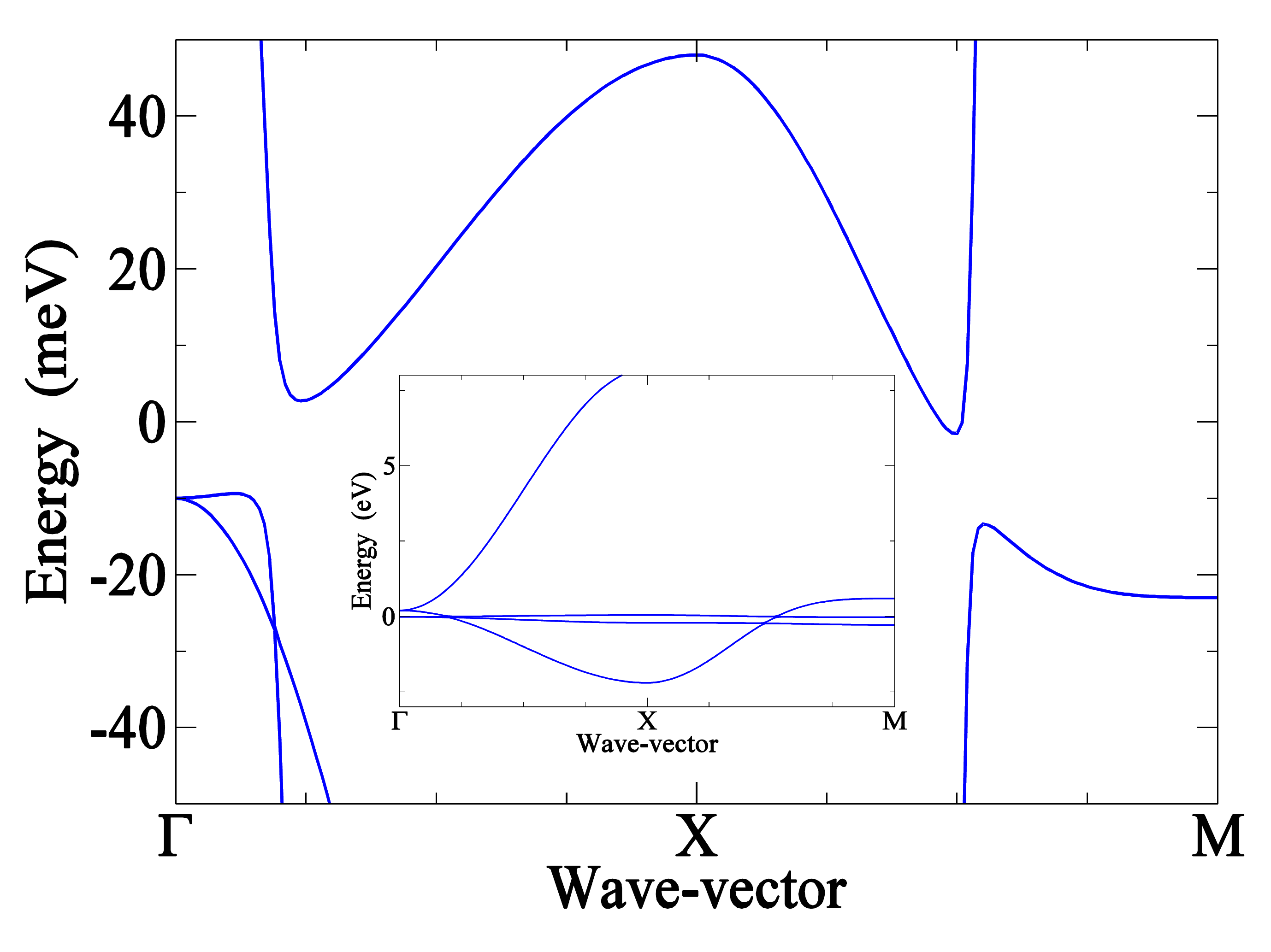}
\caption{(Color online) Bulk bandstructure, obtained by diagonalizing $H_{bulk}(\bk)$ in Eq. (\ref{Hbulk}), as a function of $\bk$ from the points $\Gamma$ to $X$ to $M$ (along $k_y=0$), showing a strongly dispersing (i.e. nearly vertical) $d$ band and a relatively flat $f$ band (i.e., nearly horizontal), with an approximate gap of $15$ meV, for the chosen values of the parameters quoted in Eq. (\ref{smb6para}). Inset: same bulk band structure, but shown over a larger window of energy. 
}\label{Fig2}
\end{figure}


\subsection{Orbital Structure and Cubic Symmetry}

Due to the underlying cubic symmetry of the local crystalline field environment of a samarium ion, the {\it fivefold} degenerate $d$ orbitals get split into {\it doubly degenerate} $e_g$ and {\it triply degenerate} $t_{2g}$ orbitals. The cubic environment, also splits the 
$J=5/2$ $f$ orbitals into a $\Gamma_{7}$ doublet and a $\Gamma_{8}$ quartet. Raman spectroscopy studies show that the dominant hybridization channel involves $f$-states of the $\Gamma_{8}$ quartet and the conduction $e_g$ states, $e^{-}+4f^{5} (\Gamma_{8}^{(\alpha)})\rightleftharpoons4f^{6}$ \cite{Sarrao1997}. It should be noted that the $e_g$ doublet is composed of $d_{x^2-y^2}$ and $d_{3z^2-r^2}$ orbitals, while the $\Gamma_8^{(\alpha)}$ ($\alpha=1,2$) $f$ quartet is composed of the following linear combination of orbitals: 
\beg\label{Gamma8}
|\Gamma_8^{(1)}\rangle=\sqrt{\frac{5}{6}}\left\vert\pm\frac{5}{2}\right\rangle+ \sqrt{\frac{1}{6}}\left\vert\mp\frac{3}{2}\right\rangle, \: \:
|\Gamma_8^{(2)}\rangle=\left\vert\pm\frac{1}{2}\right\rangle. 
\en

From the above symmetry analysis on the cubic crystal field driven splitting of the $d$ and the $f$ orbitals, it follows that the minimal tight-bonding model must involve the $\Gamma_{8}$ quartet of the localized $f$-states, and the $e_g$ quartet of the dispersive $d$ electrons, which besides being Kramers degenerate, are enriched by additional two fold {\it orbital degeneracy}. Ultimately, the {\it hybridization} among the $d$ and the $f$ electrons gives rise to the Kondo insulating phase. Therefore, a {\it minimal} Hamiltonian representing a three-dimensional cubic topological insulator, e.g., SmB$_6$, can essentially be described in terms of an {\it eight} component spinor, organized according to $\Psi^\top=\left[ \Psi_d, \Psi_f\right]$, where $\Psi_{d/f}$ are four-component spinors defined as
\begin{equation}\label{basis}
\Psi^\top_l=\left[l_{1\uparrow}, l_{1\downarrow}, l_{2\uparrow}, l_{2\downarrow} \right],
\end{equation}  
for $l=d,f$. Here $1, 2$ correspond to two orbitals of $d$- and $f$-electrons, and $\uparrow, \downarrow$ are two projections of spin. For the sake of notational simplicity, we use $\alpha=\uparrow,\downarrow$ for the Kramers doublet components of the $f$-multiplet as well. 

 It should be noted that here we have taken into account only the $\Gamma_8$ quartet and neglected $\Gamma_7$ doublet, whereas various recent numerical studies have considered both multiplets of the $f$ electrons \cite{xaidai-DFT}. However, we strongly believe that inclusion of the hybridization of $d$ electrons with $\Gamma_7$ doublet can only lead to some quantitative, but non-universal corrections for various quantities. In the Appendix we have demonstrated that hybridization with $\Gamma_8$ quartet is sufficient to produce a topologically nontrivial bulk insulating gap. Hence, the model we study serves the purpose of a minimal description that can succinctly capture the topologically robust features of this system, including the surface states, about which in a moment.


\subsection{Mean-field Hamiltonian}

At the mean-field level the full Hamiltonian of the describing Kondo insulators contains single particle terms as well as the Hubbard interaction term $(U_{ff})$ for the $f$-electrons. In the limit of infinitely strong Hubbard repulsion (i.e., $U_{ff}\to \infty$) the doubly occupied $f$-electron states are projected out and the corresponding projection operators are replaced with their mean-field values, which are then determined self-consistently. As a result, within the mean-field approximation, the effective Hamiltonian is defined through the following three terms: the hopping elements for the conduction $d$-electrons and $f$-quasiparticles, and the {\it hybridization} between these two species, however with the renormalized hopping and hybridization amplitudes \cite{CTKI}. Therefore, the eight-dimensional effective bulk Hamiltonian describing cubic topological Kondo insulators conforms to the generic form
\begin{equation}\label{Hbulk}
H_{bulk}(\bk)=\left(  \begin{array}{c c}
H^d(\bk) & V_h(\bk) \\
V^\dagger_h(\bk) & H^f(\bk)
\end{array} \right),
\end{equation} 
where $H^d(\bk)$, $H^f(\bk)$ and $V_h(\bk)$ are 4-dimensional matrices. For $l=d$ and $f$, $H^l(\bk)$ is given by 
\begin{equation}\label{tightbinding}
H^l(\bk)=\epsilon^l \hat{{\mathrm I}}_4+ t^l \left( \begin{array}{c c}
\hat{\phi}_1(\bk)+\eta_l \hat{\phi}_2(\bk) & (1-\eta_l) \hat{\phi}_3 (\bk) \\
(1-\eta_l) \hat{\phi}_3 (\bk) &  \eta_l \hat{\phi}_1(\bk)+ \hat{\phi}_2(\bk)
\end{array} 
\right),
\end{equation}
where $t^{d}$ and $t^{f}$ are the hopping amplitudes, and $\epsilon^{d}$ and $\epsilon^{f}$ are the corresponding chemical potentials, for the $d$ and $f$ electrons, respectively. In the above equation, $\hat{\phi}_j(\bk)=\hat{\sigma}_0 \phi_j(\bk)$, where $\hat{\sigma}_0$ and $\hat{{\mathrm I}}_4=\hat{\sigma}_0\otimes\hat{\sigma}_0$ are respectively the two- and four-dimensional identity matrices. Different components of the $\phi (\bk)$ functions are 
\begin{equation}\label{phis}
\begin{split}
&\phi_1(\bk)=\frac{1}{2} \left( c_x + c_y + 4 c_z \right), \\
&\phi_2(\bk)=\frac{3}{2} \left( c_x+c_y\right), \: \: \phi_3(\bk)=\frac{\sqrt{3}}{2} (c_x-c_y) \\
\end{split}
\end{equation}
with $c_\alpha=\cos{k_\alpha}$, for $\alpha=x,y,z$ (in what follows next we choose the units in which the lattice spacing $a=1$). The hybridization matrix reads as 
\begin{equation}
V_h(\bk)= \frac{V}{4}\left( \begin{array}{c c}
3 (\bar{\sigma}_x-\bar{\sigma}_y) & \sqrt{3} (\bar{\sigma}_x + \bar{\sigma}_y) \\
\sqrt{3} (\bar{\sigma}_x+\bar{\sigma}_y) & \bar{\sigma}_x-\bar{\sigma}_y+4 \bar{\sigma}_z
\end{array} 
\right),
\end{equation}
where $\bar{\sigma}_\alpha=\hat{\sigma}_\alpha \sin{k_\alpha}$ for $\alpha=x,y,z$, and $\hat{\sigma}_{x,y,z}$ are the standard two-dimensional Pauli matrices. The bare hybridization amplitude is represented by $V$. In this work we restrict ourselves with hole-like $f$ states, i.e., $t^d t^f<0$, only for which a topologically non-trivial insulating state emerges below the Kondo transition temperatures, which for SmB$_6$ is $\sim 50$ K \cite{Theory1,Sigrist,DMFT}. The resulting band structure in the bulk is shown in Fig. \ref{Fig2}, with the following choice of various parameters 
\begin{eqnarray}\label{smb6para}
&& t^d=2 \;\mbox{eV}, \: t^f=-0.05 \;\mbox{eV}, \: V= 0.0365 \; \mbox{eV}, \: \eta_d=-0.3, \nonumber \\
&& \eta_f=-0.29, \: \epsilon^{d}= 0.2 \; \mbox{eV} -3 t^d(1+\eta_d), \nonumber \\
&& \epsilon^{f}=-0.01 \; \mbox{eV} -3 t^f(1+\eta_f),  
\end{eqnarray}
appearing in $H_{bulk}(\bk)$. Interestingly, with such choice of the parameters, the bulk Kondo insulating gap is $\sim 15$ meV, resembling in this regard the observed bulk gap in SmB$_6$ in various ARPES measurements \cite{ARPES2,ARPES3,ARPES4,ARPES5,ARPES6}.

The above eight-dimensional Hamiltonian for the cubic topological Kondo insulators, $H_{bulk}(\bk)$, should be contrasted with the model Hamiltonian for weakly interacting strong $Z_2$ topological insulators [$H_{TI} (\bk)$], such as Bi$_2$Se$_3$, which, on the other hand, is {\it four-dimensional}. In the low-energy and long wavelength limit $H_{TI}(\bk)$ takes the form \cite{Zhang2}
\begin{eqnarray}\label{modelTI}
H_{TI}(\bk) &=& (A+D k^2) \;  (\hat{\tau}_0 \otimes \hat{\sigma}_0) + (M- B k^2) \; (\hat{\tau}_3 \otimes \hat{\sigma}_0) \nonumber \\
&+& V_F \vec{k} \cdot (\hat{\tau}_1 \otimes \hat{\vec{\sigma}}),  
\end{eqnarray} 
where $V_F$ is the Fermi velocity. The second term represents a parity odd but time-reversal even, inverted-band (when $M B>0$) Dirac mass. The first term gives rise to particle-hole anisotropy, and the last term yields Dirac kinetic energy in three dimensions. Here we have neglected the anisotropy among the Fermi velocities along different directions, arising from the underlying crystalline structure \cite{Zhang2}. Two sets of Pauli matrices, $\tau$ and $\sigma$, respectively, operate on the even-odd parity band and the spin index. Next we argue although  $H_{bulk}(\bk)$ is eight-dimensional, it still represents a three dimensional $Z_2$ topological insulators, however, generalized for {\it multi-band} systems. To perform this exercise we first need to reorganize the spinor basis according to 
\begin{equation}\label{spinorredefined}
\Psi^\top_l =\left[l_{1\uparrow}, l_{2\uparrow}, l_{1\downarrow}, l_{2\downarrow} \right],
\end{equation}
for $l=d,f$ and define $\Psi^\top=[\Psi_d, \Psi_f]$. This reorganization is tantamount of a unitary transformation that exchanges second and third entries, and also sixth and seventh entries in $H_{bulk}(\bk)$. In the unitarily rotated spinor basis
\begin{equation}\label{bulksmB6}
\begin{split}
& H_{bulk}(\bk)=\left(\hat{\tau}_0 \otimes \hat{\sigma}_0 \right)\otimes \hat{H}_+ + \left( \hat{\tau}_3 \otimes \hat{\sigma}_0 \right)\otimes \hat{H}_- \\
&+ \left( \hat{\tau}_1 \otimes \hat{\sigma}_3 \right) \otimes \hat{V}_z + \left(\hat{\tau}_1 \otimes \hat{\sigma}_1 \right) \otimes \hat{V}_x+ \left( \hat{\tau}_1 \otimes \hat{\sigma}_2 \right) \otimes \hat{V}_y. \\
\end{split}
\end{equation}   
The Pauli matrices $\hat{\tau}_j$ operate on $(d,f)$ states, while $\hat{\sigma}_j$ operate in spin space, and $\hat{H}_{\pm},\hat{V}_\alpha$ operate in orbital subspace, spanned by $l_1$ and $l_2$ for $l=d,f$. The orbital components of various matrices go as
\beg
\begin{split}
&\hat{V}_x=\frac{V}{4}\left( \begin{matrix}
3 & \sqrt{3} \\
\sqrt{3} & 1
\end{matrix}
\right) \sin{k_x},\\
&\hat{V}_y=\frac{V}{4}\left( \begin{matrix}
-3 & \sqrt{3} \\
\sqrt{3} & -1
\end{matrix}
\right) \sin{k_y}, 
~\hat{V}_z=\frac{V}{4}\left( \begin{matrix}
0 & 0 \\
0 & 4
\end{matrix}
\right) \sin{k_z},
\end{split}
\en
and $\hat{H}_{\pm}=\frac{1}{2} \left[ \hat{H}_d (\bk) \pm \hat{H}_f (\bk) \right]$, where for $l=d,f$
\begin{eqnarray}
\hspace{-3cm} \hat{H}_l(\bk) = \quad \quad \quad \quad \quad \nonumber \\
\left( \begin{array} {c c}
\epsilon^l + t^l \phi_1(\bk) + t^l \eta_l \phi_2 (\bk) & t^l (1-\eta_l) \phi_3 (\bk) \\
t^l (1-\eta_l) \phi_3 (\bk) &  \epsilon^l + t^l \phi_2(\bk) + t^l \eta_l \phi_1 (\bk)
\end{array}
\right).
\end{eqnarray}

The following identification of various terms appearing in Eq. (\ref{bulksmB6}), in conjunction with its comparison with Eq. (\ref{modelTI}), allows us to conclude that $H_{bulk}(\bk)$ represents a multi-band strong $Z_2$ topological insulator in three dimensions: terms proportional to $\hat{V}_\alpha$ define Dirac kinetic energy in three dimensions, $\hat{H}_-$ represents time-reversal symmetric, odd-parity, inverted-band Dirac mass, and $\hat{H}_+$ gives rise to particle-hole asymmetry. Equivalent quantities in $H_{TI}(\bk)$ are replaced by {\it scalar} entries. The Parity operator in this basis reads as $\hat{P}=\hat{\tau_3} \otimes \sigma_0 \otimes I_2$, where $I_2$ is a two-dimensional unit matrix, here operating on the orbital subspace. It should be noted that $H_{bulk} (\bk)$ describes strong $Z_2$ topological insulator only when $H_-$ is not diagonal, which is satisfied for any $\eta_{d,f} \neq 1$. Hence, $H_{bulk} (\bk)$ can be generalized for multi-band strong $Z_2$ topological insulators, where the {\it dimensionality} of $\hat{V}_\alpha$, $\hat{H}_-$ and $\hat{H}_+$ corresponds to the number of orbitals participating in the low energy dynamics. To further substantiate our claim, we also compute the topological invariant with the above model, shown in the Appendix, confirming that $H_{bulk}(\bk)$ represents a strong $Z_2$ topological insulator in three dimensions.

In the bulk Hamiltonian $H_{bulk}(\bk)$, we can add a term ${\cal M}=\Delta_{PT} \: \left( \hat{\tau}_2 \otimes \hat{\sigma}_0 \otimes \hat{I}_2 \right)$, representing a parity and time-reversal odd Dirac mass, which anticommutes with $H_{bulk}(\bk)$, i.e. $\left\{ H_{bulk}(\bk), {\cal M} \right\}=0$. Therefore, together $H_{bulk}(\bk)+ {\cal M}$ represents an {\it axionic} state of matter. The time-reversal operator in our representation reads as $I_T=\left( \hat{\tau_0} \otimes \hat{\sigma_2} \otimes \hat{I}_2 \right) \; K$, where $K$ is the complex conjugation. Recently, axionic ground state has been proposed for various magnetic topological insulators \cite{axion1, axion2, axion3}, as well as for paired ground state in various three dimensional narrow gap semiconductors with $p+is$ pairing symmetries \cite{pisaxion}. On the other hand, in the present situation the parity and time-reversal odd Dirac mass corresponds to a Kondo singlet state \cite{goswami-roy-Kondo}, and can, in principle, be favored by strong interactions between the conduction $d$ and localized $f$ electrons.  


\section{Surface states}
\label{surface}

The nontrivial $Z_2$ topological invariant of the bulk makes topological insulators distinct from a trivial vacuum, and therefore an interface between these two systems hosts topologically protected metallic surface states. Next we proceed to find the low-energy Hamiltonian for such surface states. Let us first outline the strategy of finding the surface Hamiltonian. Without any loss of generality, we will only consider surfaces that are perpendicular to the main cubic axes in this paper. For definiteness, we focus on the (001) surface on which the momentum components $k_x$ and $k_y$ remain good quantum numbers.

Here we assume that the even ($d$ electron) and odd ($f$ electron) parity bands invert at one of the high-symmetry points of the BZ, denoted by $\bk_m$. To determine the energy $E_m$ of the electrons at the Dirac point, we expand $H_{bulk} (\bk)$ up to the second order in $\delta\bk=\bk-\bk_m$ and then set $\delta k_x=\delta k_y=0$, while replacing $\delta k_z\to-i\partial_z$. The energy $E_m$ is then an eigenvalue of the Schr\"{o}dinger equation
\beg\label{EmDirac}
H_{bulk}(\delta k_z\to-i\partial_z)\Psi(z)=E_m\Psi(z).
\en
We here consider a semi-infinite sample, occupying the region $z>0$, with a {\it sharp} boundary at $z=0$ and vacuum for $z<0$. Therefore, the wave function of the surface bound state $\Psi(z)\propto e^{-\lambda z}$, where $\lambda$ corresponds to the penetration depth of the surface states into the bulk. One of the boundary conditions $\Psi(z \to \infty)=0$, imposes a constraint over $\lambda$, $\Re(\lambda)>0$. The effective surface Hamiltonian will then be obtained by averaging out $H_{bulk}$ evaluated at finite $\delta k_{x,y}$ over $\Psi(z)$:\cite{Zhang1, Zhang2, Surface1}
\beg\label{Hsurf}
H_{surf}(k_x,k_y)=\int\limits_0^\infty dz\langle \Psi(z)|H_{bulk}(k_x,k_y;z)|\Psi(z)\rangle.
\en
Below, we subscribe the above methodology to derive the surface Hamiltonian for the family of cubic topological Kondo insulators, such as SmB$_6$, with the bulk Hamiltonian shown in Eq. (\ref{bulksmB6}).


\subsection{Effective Hamiltonian Near $\overline{Y}=(0, \pi)$ Point}

To obtain the effective Hamiltonian near the $\bar{Y}$ point of the surface BZ we need to expand $H_{bulk}(\bk)$ around $(0,\pi,0)$ point. In the vicinity of $(0,\pi,0)$ point various functions appearing in $H_{bulk}(\bk)$ to the leading order are
\beg
\begin{split}
&\phi_1(\bk) \rightarrow 2-2k^2_z, ~\phi_2 (\bk) \rightarrow 0, ~\phi_3 (\bk) \rightarrow \sqrt{3}, \\
&\sin{k_z}\rightarrow k_z, ~\sin{k_x} \rightarrow k_x, ~\sin{k_y} \rightarrow -k_y.
\end{split}
\en
For the calculation of surface states, we, once again, need to organize the spinor basis slightly different than in Eq. (\ref{spinorredefined}). For convenience, let us define the eight-component spinor as $\Psi^\top=\left[ \Psi_\uparrow, \Psi_\downarrow \right]$, where $\Psi^\top_\sigma= \left[ d_{1\sigma}, d_{2\sigma}, f_{1\sigma}, f_{2\sigma} \right]$, for $\sigma=\uparrow, \downarrow$. In this basis the eight-dimensional Hamiltonian $H_{bulk} (\bk)$ becomes 
\begin{eqnarray}\label{bulkHamilsurfstate}
H_{bulk} (\bk) &=& \left( \begin{array} {c c | c c}
\hat{H}_- & \hat{V}_z & \tilde{0}_2 & \hat{V}_x-i \hat{V}_y \\
\hat{V}_z & -\hat{H}_- & \hat{V}_x-i \hat{V}_y & \tilde{0}_2 \\
\hline
\tilde{0}_2 & \hat{V}_x+i \hat{V}_y & \hat{H}_- & -\hat{V}_z \\
\hat{V}_x+i \hat{V}_y & \tilde{0}_2 & -\hat{V}_z & -\hat{H}_-
\end{array}
\right) \nonumber \\
&+& \left( \begin{array} {c c | c c}
\hat{H}_+ & \tilde{0}_2 & \tilde{0}_2 & \tilde{0}_2 \\
\tilde{0}_2 & \hat{H}_+ & \tilde{0}_2 & \tilde{0}_2 \\
\hline
\tilde{0}_2 & \tilde{0}_2 & \hat{H}_+ & \tilde{0}_2 \\
\tilde{0}_2 & \tilde{0}_2 & \tilde{0}_2 & \hat{H}_+
\end{array}
\right)
\equiv \left( \begin{array} {c|c}
H_{\uparrow \uparrow} & H_{\uparrow \downarrow} \\
\hline
H^\dagger_{\uparrow \downarrow} & H_{\downarrow \downarrow}
\end{array}
\right),
\end{eqnarray}
where $\tilde{0}_2$ represents two-dimensional {\it null matrix}, and $H_{\uparrow \uparrow}$, $H_{\downarrow \downarrow}$, $H_{\uparrow \downarrow}$ are $4 \times 4$ matrices. For the calculation of surface bound states we first set $\hat{V}_x, \hat{V}_y=\tilde{0}_2$. After obtaining the solutions of the surface states, say $| \Psi_\uparrow\rangle$ and $| \Psi_\downarrow \rangle$, the eigenstates of $H_{\uparrow \uparrow}$ and $H_{\downarrow \downarrow}$, respectively, we will perform a perturbative expansion of $H_{\uparrow \downarrow}$, $H^\dagger_{\uparrow \downarrow}$, in the two-dimensional basis spaced by $| \Psi_\uparrow\rangle$ and $| \Psi_\downarrow \rangle$ to obtain the surface Hamiltonian.  

\begin{figure}[htb]
\includegraphics[width=8.25cm,height=6.0cm]{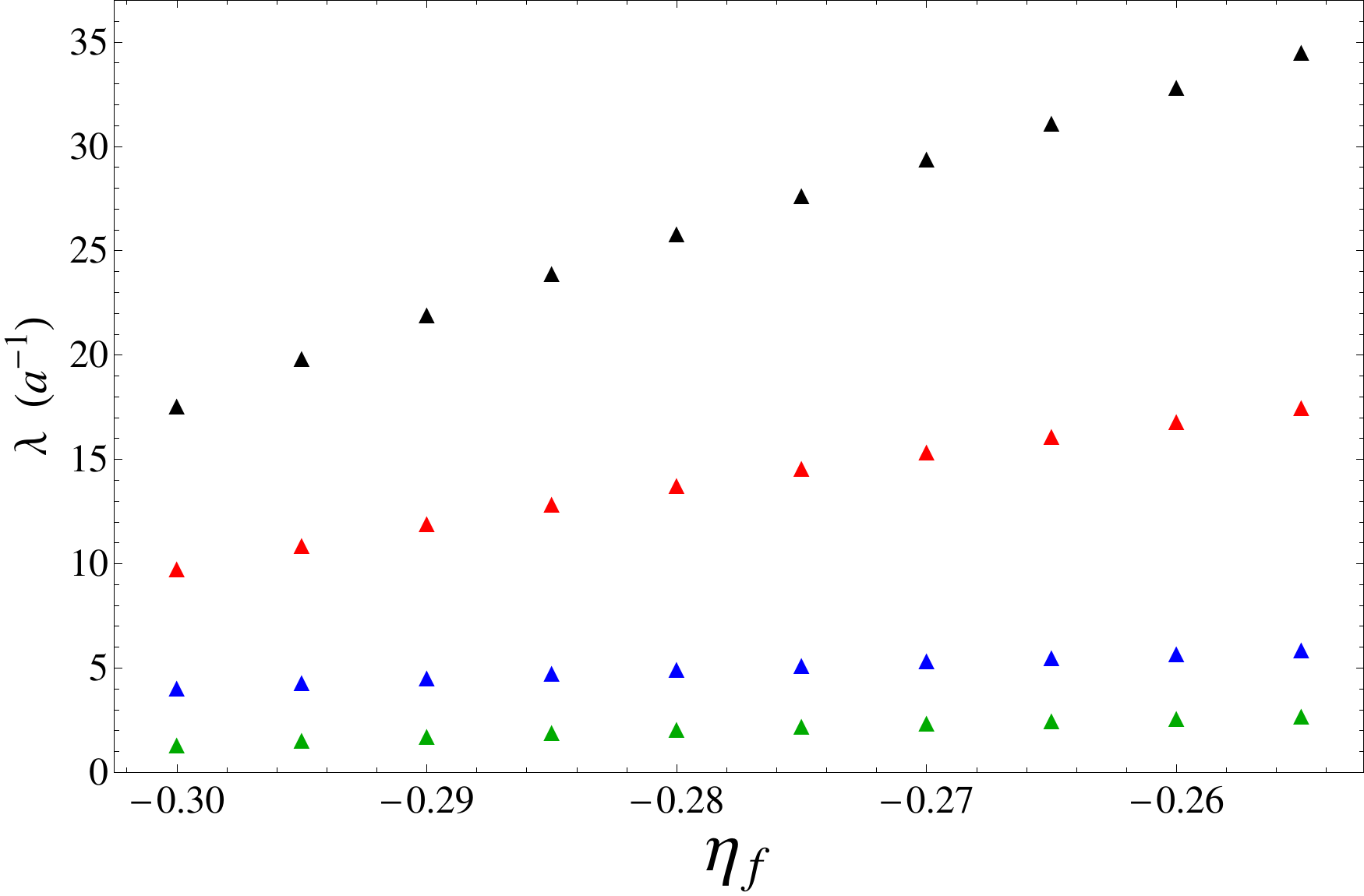}
\caption{(Color online) Dependence of four roots of $\lambda$ [solutions of Eq. (\ref{secularY})], for fixed value of the parameter $\eta_d=-0.3$, but as a function of $\eta_f$, appearing in Eq. (\ref{tightbinding}). Rest of the parameters of $H_{bulk}(\bk)$ are the same as in Eq. (\ref{smb6para}). Dependence of the location of the Dirac points ($E_{\bar{X}}$) on this parameter is shown in Fig. 4. The smallest root of $\lambda$ determines the penetration depth of the surface state into the bulk.}
\end{figure}

Next we make an ansatz for the surface states (dropping the spin index in $| \Psi_{\uparrow, \downarrow} \rangle$ from now on for the sake of notational simplicity) $\Psi(z) \sim \exp{(-\lambda z)} \Psi(\lambda)$. Taking $k_z \rightarrow -i \partial_z$, we here first wish to solve
\begin{equation}
H_{\uparrow \uparrow} (k_z \rightarrow -i \lambda)\Psi(\lambda)= E_{\bar{Y}} \; \Psi(\lambda),
\end{equation}
where $E_{\bar{Y}}$ is the energy of the surface states at $\bar{Y}=(0,\pi)$ point of the surface BZ. The above equation introduces a set of constraints among various spinor components as follow
\begin{eqnarray}
d_2 (\lambda)= G_d (\lambda) d_1(\lambda) \quad \mbox{and} \quad f_2(\lambda) = G_f (\lambda) f_1 (\lambda),
\end{eqnarray}
where 
\beg
\begin{split}
&G_l(\lambda)=-\left( \frac{\epsilon^l_1+ t^l \lambda^2-E_{\bar{Y}}}{\epsilon^l_3} \right), \\
&\epsilon^l_1=\epsilon^l+ 2 t^l, ~\epsilon^l_3=\sqrt{3} t^l (1-\eta_l),
\end{split}
\en
for $l=d,f$. The remaining two spinor components are related according to $f_1 (\lambda)= H(\lambda) d_1 (\lambda)$, where 
\begin{equation}
H(\lambda)= \frac{i V\lambda G_d(\lambda)}{\left( \epsilon^f_2+ t^f \eta_f \lambda^2 -E_{\bar{Y}} \right) G_f(\lambda) -\epsilon^f_3},
\end{equation}
and $\epsilon^l_2= \epsilon^l + 2 \eta_l t^l$. A nontrivial solution of all the spinor components yields the secular equation
\begin{equation}\label{secularlambda}
\tilde{V}^2 \lambda^2 + \left[ \lambda^2 + \tilde{\Lambda}^2_d -\frac{\Pi^2_d}{\lambda^2 + \Lambda^2_d} \right]
\left[ \lambda^2 + \tilde{\Lambda}^2_f -\frac{\Pi^2_f}{\lambda^2 + \Lambda^2_f} \right]=0, 
\end{equation}
where
\begin{eqnarray}
\begin{split}
&\tilde{V}^2=\frac{V^2}{t^d t^f \eta_d \eta_f}, \:
\Lambda^2_l=\frac{\epsilon^l_1-E_{\bar{Y}}}{t^l}, \:
\tilde{\Lambda}^2_l=\frac{\epsilon^l_2-E_{\bar{Y}}}{t^l \eta_l}, \\
&\Pi^2_l=\frac{3 (1-\eta_l)^2}{\eta_l}, \\
\end{split}
\end{eqnarray}
for $l=d,f$.

The above equation altogether yields eight roots of the form $\pm \lambda_j$, for $j=1,2,3,4$. Upon imposing the boundary condition $\Psi(z \rightarrow \infty)=0$, the surface state gets restricted to the following form 
\begin{eqnarray}\label{surfacegeneral}
\Psi(z)=\sum_{j=1,2,3,4} C_j \exp(-\lambda_j z) \Psi(\lambda_j),
\end{eqnarray}  
where $C_j$s are arbitrary constant, which can now be eliminated from the second boundary condition $\Psi(z=0)=0$. Here we have assumed that $\mbox{Re}(\lambda)_j>0$, for $j=1,2,3,4$. This assumption is justified, since all the coefficients in Eq. (\ref{secularlambda}) are real. Upon imposing the above boundary condition we obtain the following algebraic equation:
\begin{equation}\label{secularY}
G^1_f H_1 -X G^2_f H_2 - Y G^3_f H_3 - Z G^4_f H_4=0, 
\end{equation}
from which one can immediately determine the energy $E_{\bar{Y}}$. The above equation is too complicated to obtain a closed analytic expression for $E_{\bar{Y}}$. We here obtain its solution numerically. Scaling of four roots of $\lambda$ as a function of the parameter $\eta_f$, while keeping the rest of the parameters in $H_{bulk}(\bk)$ fixed at their values, quoted in Eq. (\ref{smb6para}), is shown in Fig.~3. Various quantities appearing in the last equation are 
\begin{widetext}
\beg
X=\frac{H_1-H_{3,4} G^{1,3,4}_d}{H_2- H_{3,4} G^{2,3,4}_d},
Y=\frac{1}{G^3_d-G^4_d} \left[ \left( G^1_d-G^4_d \right)-\left( G^2_d-G^4_d \right) X \right],
Z=1- X - \left\{ \left( \frac{G^1_d-G^4_d}{G^3_d-G^4_d}\right) -X \left( \frac{G^2_d-G^4_d}{G^3_d-G^4_d}\right) \right\},
\en
where
\beg
G^j_l \equiv G_l (\lambda_j), ~H_j \equiv H(\lambda_j), 
H_{3,4} G^{k,3,4}_d = H_4 + \left( H_3-H_4\right) \left(\frac{ G^k_d-G^4_d}{G^3_d-G^4_d}\right),
\en
\end{widetext}
 $l=d,f$, $j=1,2,3,4$ and $k=1,2$. Arbitrary coefficients appearing in Eq. (\ref{surfacegeneral}) are related to the above parameters according to
\beg 
\frac{C_2}{C_1} =-X  \left( \frac{d^1_1}{d^2_1} \right),  \frac{C_3}{C_4}=-Y  \left( \frac{d^1_1}{d^3_1} \right), 
\frac{C_4}{C_1}=-Z  \left(\frac{d^1_1}{d^4_1}\right), 
\en
where $l^j_k=l_k(\lambda_j)$ for $l=d,f$, $j=1,2,3,4$, and $k=1,2$. In terms of these new parameters the surface state is 
\begin{equation}\label{Ywaveup}
| \Psi_\uparrow (z) \rangle= C_1 \sum_{q=1,2,3,4} {\cal X}_q \: \exp{\left( -\lambda_q z \right)} \:  
\left[ \begin{array}{c}
1 \\ G^q_d \\ H_q \\ H_q G^q_f
\end{array} 
\right],
\end{equation}
where ${\cal X}_1=1$, ${\cal X}_2=-X$, ${\cal X}_3=-Y$, ${\cal X}_4=-Z$. Here we have reintroduced the spin-index in the wave function. The remaining arbitrary constant $C_1$ determines the overall normalization factor of $| \Psi_\uparrow(z) \rangle$. After some lengthy but straightforward calculation it can be shown that the other surface bound state $| \Psi_\downarrow (z) \rangle$, satisfying 
\begin{equation}
H_{\downarrow \downarrow} (k_z \to i \lambda) \Psi_\downarrow(\lambda)=E_{\bar{Y}} \Psi_\downarrow (\lambda),
\end{equation}  
is identical to $| \Psi_\uparrow (z) \rangle$, shown in Eq. (\ref{Ywaveup}). From the numerical solution of the wave-functions $|\Psi_{\uparrow/\downarrow} (z) \rangle$, we find that magnitudes of all the four components of the spinor wave functions are comparable with each other.

Next we perform the perturbative expansion of the off-diagonal components of $H_{bulk}$ in Eq. (\ref{bulkHamilsurfstate}), yielding the surface Dirac Hamiltonian at $Y=(0,\pi)$ point of the surface BZ
\begin{equation}\label{YHamil}
\begin{split}
H^{\bar{Y}}_{sur}= & \int^{\infty}_0 dz \left[ \begin{array}{c c}
0 & \langle \psi_\uparrow (z) | H_{\uparrow \downarrow}| \Psi_\downarrow (z) \rangle \\
\langle \psi_\downarrow (z) | H^\dagger_{\uparrow \downarrow}| \Psi_\uparrow (z) \rangle & 0
\end{array} 
\right], \\ 
&=(v_x^{\bar{Y}}\sigma_x k_x -v_y^{\bar{Y}} \sigma_y k_y).
\end{split}
\end{equation}
In the above Hamiltonian $v_x^{\bar{Y}} \neq v_y^{\bar{Y}}$, and thus $H^{\bar{Y}}_{sur}$ describes an {\it anisotropic} Dirac cone at $\overline{Y}$ point. However, due to the complex nature of the algebraic equation [Eq. (\ref{secularY})], expressions for $v_{x,y}^{\bar{Y}}$ and $\lambda$s are quite lengthy and they cannot be expressed compactly. We, therefore, perform numerical diagonalization to obtain the surface band structure (see Fig.~4) that captures the essential properties of the surface states.


\subsection{Effective Hamiltonian Near $\overline{X}=(\pi,0)$ Point}

To arrive at the effective Hamiltonian for the surface states near the $\bar{X}$ point, we need to expand $H_{bulk}(\bk)$ around $(\pi,0,0)$, yielding
\begin{equation}
\begin{split}
&\phi_1(\bk)\rightarrow 2-2 k^2_z, \; \phi_2(\bk) \to 0, \; \phi_3(\bk) \to -\sqrt{3}, \\
&\sin{k_z} \to k_z, \; \sin{k_x} \to -k_x,\; \sin{k_y} \to k_y.
\end{split}
\end{equation}
Otherwise, the calculation of the surface states near $(\pi,0,0)$ are exactly the same as the one near $(0,\pi,0)$ point, shown in previous subsection. The surface Hamiltonian in the vicinity of the $\overline{X}=(\pi,0)$ point reads as 
\begin{equation}\label{XHamil}
H^{\bar{X}}_{sur} \;=\: (v_x^{\bar{X}} \sigma_x k_x-v_y^{\bar{X}} \sigma_y k_y),
\end{equation}
and once again $v_x^{\bar{X}}\neq v_y^{\bar{X}}$. Therefore, $H^{\bar{X}}_{sur}$ also represents an {\it anisotropic} Dirac cone near the $\bar{X}$ point of the surface BZ. We also notice that $v_x^{\bar{X}}=v_y^{\bar{Y}}$ and $v_y^{\bar{X}}=v_x^{\bar{Y}}$, reflecting a fourfold $C_4$ rotational symmetry on the surface, resulting from the underlying cubic symmetry in the bulk, which has been confirmed in recent measurement of magnetoresistance \cite{oldosi-1, oldosi-2}. The location of the Dirac fermions near $\bar{X}$ and $\bar{Y}$ points are also the same, i.e., $E_{\bar{X}}=E_{\bar{Y}}$. From now on we will refer $\bar{X}$ and $\bar{Y}$ points of the surface BZ together as $\bar{X}$ points.


\subsection{Effective Hamiltonian Near $\overline{\Gamma}=(0,0)$ Point}

Next we proceed to find the surface state and the corresponding Hamiltonian near the $\bar{\Gamma}=(0,0)$ point of the surface BZ. In this case we can obtain analytical expression for both penetration depth ($\lambda$) and Fermi velocity ($v_F$) of the surface states. In the vicinity of $(0,0,\pi)$ point various function appearing in $H_{bulk} (\bk)$ are 
\begin{equation}
\begin{split}
&\phi_1(\bk)\rightarrow -1+ k^2_z, ~\phi_2(\bk) \to 3, ~\phi_3(\bk) \to 0, \\
&\sin{k_z} \to -k_z, \; \sin{k_x} \to k_x,\; \sin{k_y} \to k_y.
\end{split}
\end{equation} 
Once again we can bring the bulk Hamiltonian in the form as in Eq. (\ref{bulkHamilsurfstate}) to calculate the surface bound states and surface Hamiltonian, and solve for $H_{\uparrow \uparrow} (k_z \to -i \lambda) \Psi_\uparrow (\lambda)=E_{\bar{\Gamma}} \Psi_\uparrow(\lambda)$. In the vicinity of $(0,0,\pi)$ point this equation simplifies significantly, immediately yielding (once again here we are dropping the spin index from the spinor components for notational simplicity)
\begin{equation}
d_1 (\lambda)=f_1 (\lambda)=0.
\end{equation}  
The rest of the components satisfy 
\begin{equation}
\begin{split}
&(A^d-\lambda^2) d_2(\lambda)+i \tilde{V}_d \lambda f_2 (\lambda)=0, \\
&(A^f-\lambda^2) f_2(\lambda)+i \tilde{V}_f \lambda d_2 (\lambda)=0,
\end{split}
\end{equation}
where 
\begin{equation}
\begin{split}
&A^l=\frac{\epsilon^l_2 - E_{\bar{\Gamma}}}{\eta_l t^l}, \quad \tilde{V}_l=\frac{V}{\eta_l t_l}, \\
&\epsilon^l_1=\epsilon^l + t^l (3 \eta_l-1), \quad \epsilon^l_2=\epsilon^l + t^l (3- \eta_l), 
\end{split}
\end{equation}
for $l=d,f$. Notice that $\epsilon^l_{1,2}$ are slightly different near $(0,0,\pi)$ and $(0,\pi,0)$, although to avoid notational complication, we are using the same symbols. Nontrivial solutions of the spinor components, yield four roots of $\lambda$, of the form $\pm \lambda_j$, and for $j=1,2$ we have
\begin{equation}
\begin{split}
\lambda_j=&\frac{1}{\sqrt{2}} \bigg[ \left( A^d+A^f \right)- \tilde{V}_d \tilde{V}_f \\ 
& + (-1)^j \sqrt{\left( A^d+A^f- \tilde{V}_d \tilde{V}_f\right)^2-4 A^d A^f } \; \bigg]^{1/2}.
\end{split}
\end{equation}

\begin{figure}[htb]
\includegraphics[width=8.25cm,height=6.0cm]{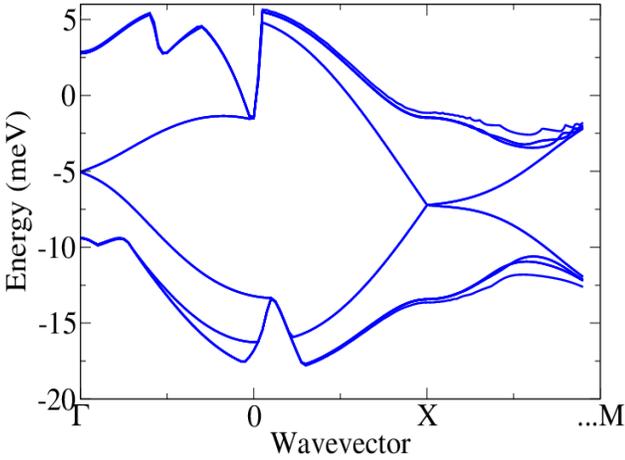}
\caption{(Color online) Surface band structure plotted along the same directions (but not  all the way to $M$) shows an isotropic Dirac point at $\Gamma$ and a strongly anisotropic Dirac point at $X$. The relative shift of the Dirac points at $X$ and $\Gamma$ are controlled by the parameters $\eta_d$ and $\eta_f$, which are tuned so that the Dirac point at $X$ is below $\Gamma$. The anisotropy (i.e., ratio of the velocities $v_x/v_y$) of the Dirac cone at $X$ varies from $v_x/v_y=20$ in the vicinity of the  Dirac point to $v_x/v_y=3$ at energies away from the Dirac point.
}\label{Fig4}
\end{figure}

Imposing boundary condition $\Psi_\uparrow (z \rightarrow \infty)= 0$, we can write the surface bound state as 
\begin{equation}
| \Psi_\uparrow (z) \rangle= C_1 \left\{
e^{-\lambda_1 z}  \left[ \begin{array} {c}
0 \\ d^1_2 \\ 0 \\f^1_2
\end{array}
\right]
+ A \: e^{-\lambda_2 z}  \left[ \begin{array} {c}
0 \\ d^2_2 \\ 0 \\f^2_2
\end{array}
\right]
\right\}.
\end{equation}
Upon imposing the second boundary condition $\Psi_\uparrow (z=0)=0$, we obtain $\left(A^d-\lambda^2_1 \right) \lambda_2=\left(A^d-\lambda^2_2 \right) \lambda_1$, yielding 
\begin{equation}
\begin{split}
E_{\bar{\Gamma}}=&\frac{\epsilon^f_2 t^d\eta_d-\epsilon^d_2 t^f \eta_f}{t^d \eta_d-t^f \eta_f}, \\=& 
\frac{t^f \eta_f \left( \epsilon^d + t^d (3-\eta_d)\right) - t^d \eta_d \left( \epsilon^f + t^f (3-\eta_f)\right)}{t^f \eta_f-t^d \eta_d}.
\end{split}
\end{equation}
The wave function for the surface state can then be compactly written as  
\begin{equation}
| \Psi_\uparrow (z) \rangle= C_1 \left( e^{-\lambda_1 z}-e^{-\lambda_2 z} \right)
\left[ \begin{array} {c}
0 \\ d^1_2 \\ 0 \\f^1_2
\end{array}
\right],
\end{equation}
where $l^{j}_{1,2} \equiv l_{1,2} (\lambda_j)$ for $j=1,2$ and $l=d,f$, and $C_1$ stands as an overall normalization constant. Performing the similar analysis for the surface bound state for the $\downarrow$ component of the spin projection, we find that $|\Psi_\downarrow(z) \rangle=|\Psi_\uparrow(z)\rangle$. A perturbative expansion of $H_{\uparrow \downarrow}$ and $H^{\dagger}_{\uparrow \downarrow}$ in the basis of $|\Psi_\uparrow(z)\rangle$ and $|\Psi_\downarrow(z)\rangle$, yields the surface Hamiltonian in the vicinity of the $\overline{\Gamma}$ point
\begin{equation}\label{gamaHamil}
H^{\bar{\Gamma}}_{sur}=v_F^{\bar{\Gamma}} \left( \sigma_x k_x - \sigma_y k_y\right),
\end{equation} 
which represents an {\it isotropic} Dirac cone, with the Fermi velocity 
\beg
v_F^{\bar{\Gamma}}=2V\sqrt{\frac{-2t^dt^f\eta_d\eta_f}{(\eta_dt^d-\eta_ft^f)^2}}.
\en
Note that the Fermi velocity is of the order of hybridization amplitude, which implies that $v_F^{\bar{\Gamma}}\ll p_F/m$, where $m$
is a bare electron mass and $p_F$ is a Fermi momentum. From the solution of the wave functions, we find that in $| \Psi_{\uparrow, \downarrow}(z) \rangle$, $d_2 (\lambda_1) \ll f_2 (\lambda_1)$. Therefore, the overlap between the $d$ and $f$ electrons for the surface states near the $\bar{\Gamma}$ point is small, in contrast to the situation near $\bar{X}$ points.  

\begin{figure}[htb]
\includegraphics[width=8.25cm,height=6.5cm]{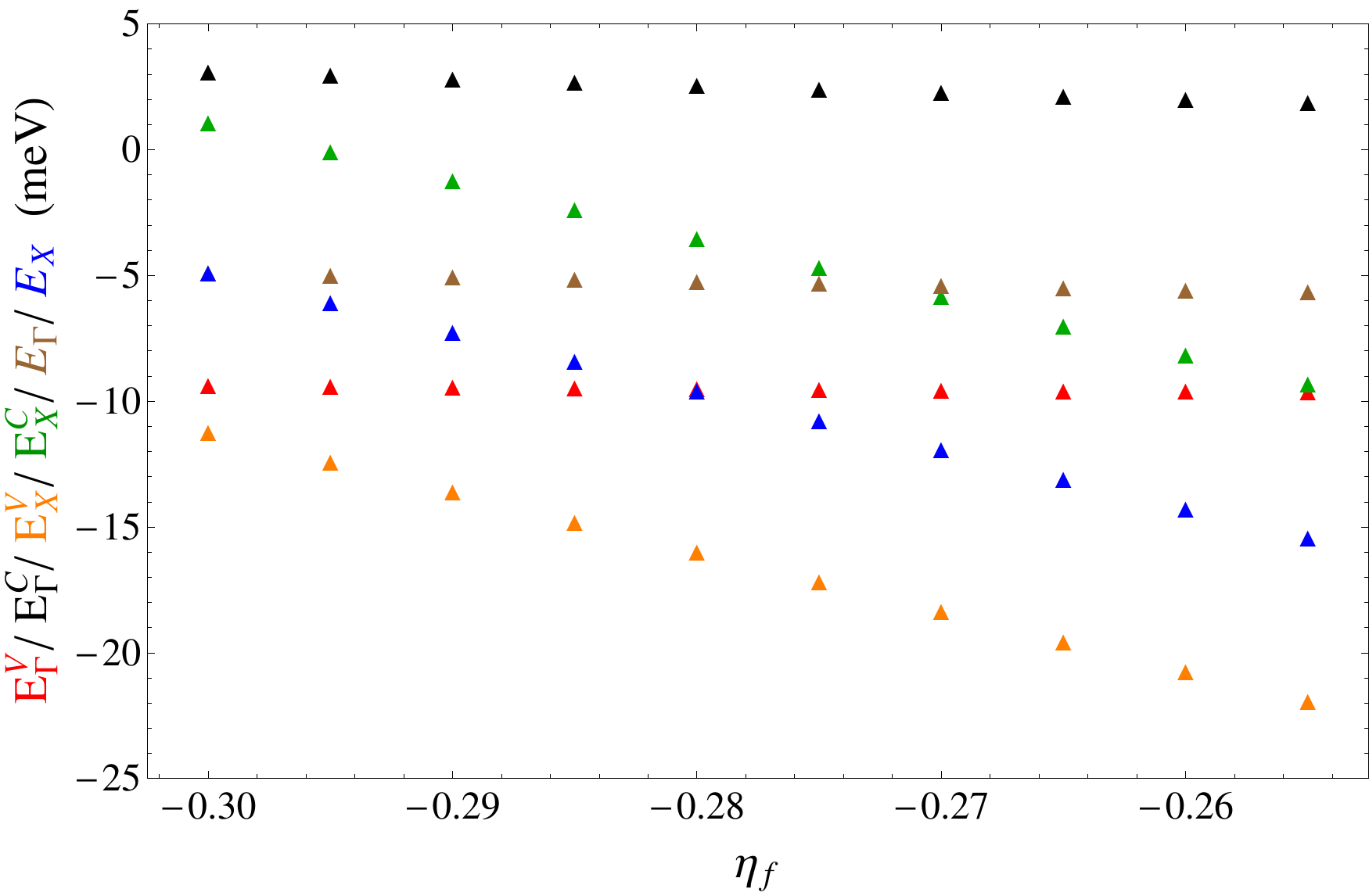}
\caption{(Color online) Dependence of various quantities in the bulk as well as on the surface on $\eta_f$, while rest of the parameters are kept fixed to their values quoted in Eq. (\ref{smb6para}). Here, $E^{V}_{\Gamma/X}$ represents the top of the valence band at $\Gamma/X$ point (shown in red/orange) and $E^{C}_{\Gamma/X}$ stands for the bottom of the conduction band at $\Gamma/X$ point of the bulk BZ (shown in black/green), of the $3D$ bulk BZ. $E_\Gamma$ (brown) and $E_X$ (blue) are the energies of the Dirac fermions near the $\bar{\Gamma}$ and $\bar{X}$ point of the surface BZ. This figure shows that at least the Dirac point at the $\bar{X}$ point can be dragged down into the valence band by tuning some parameter ($\eta_f$ for example here) in the theory.} \label{Fig5}
\end{figure}


\subsection{Surface Band Structure}

Next we numerically compute the surface state of the bulk SmB$_6$ from the model Hamiltonian $H_{bulk}(\bk)$ on the (001) surface. For this purpose we can treat momentum $k_x$ and $k_y$ to be constant, and represent the bulk Hamiltonian as $H_{bulk}(\bk) \equiv H_{bulk}(k_x,k_y,k_z)$, where
\beg
H_{bulk}(k_x,k_y,k_z)=h(k_x,k_y)+[\rho(k_x,k_y)e^{i k_z} + \; h.c],
\en  
and $h(k_x,k_y), \rho(k_x,k_y)$ are defined as follows:
\begin{eqnarray}
h(k_x,k_y)&=& \frac{H_{bulk}(k_x,k_y,0)+H_{bulk}(k_x,k_y,\frac{\pi}{a})}{2},\\
\rho(k_x,k_y) &=& \frac{H_{bulk}(k_x,k_y,0)-H_{bulk}(k_x,k_y,\frac{\pi}{a})}{4} \nonumber \\
&-& \frac{i}{2} \; H_{bulk}(k_x,k_y,\frac{\pi}{2 a}) + \frac{i}{2} \; h(k_x,k_y).
\end{eqnarray}
To compute the surface states as well as the surface band structure, we first need to Fourier transform the Hamiltonian $H_{bulk}(k_x,k_y,k_z)$ to real space along the $z$-axis yielding
\beg
\begin{split}
H_1(k_x,k_y)=&\sum_{n=0}^N [h(k_x,k_y)\otimes \ket{n}\bra{n}\\&+[\rho(k_x,k_y)\otimes \ket{n}\bra{n+1}+h.c],
\end{split}
\en  
and we set $N=180$. Upon numerically diagonalizing the above Hamiltonian $H_1 (k_x, k_y)$, we obtain the spectrum of the surface states, shown in Fig.~\ref{Fig4}, for a particular set of parameters quoted in Eq. (\ref{smb6para}). 

\begin{figure}[htb]
\includegraphics[width=8.5cm,height=7.0cm]{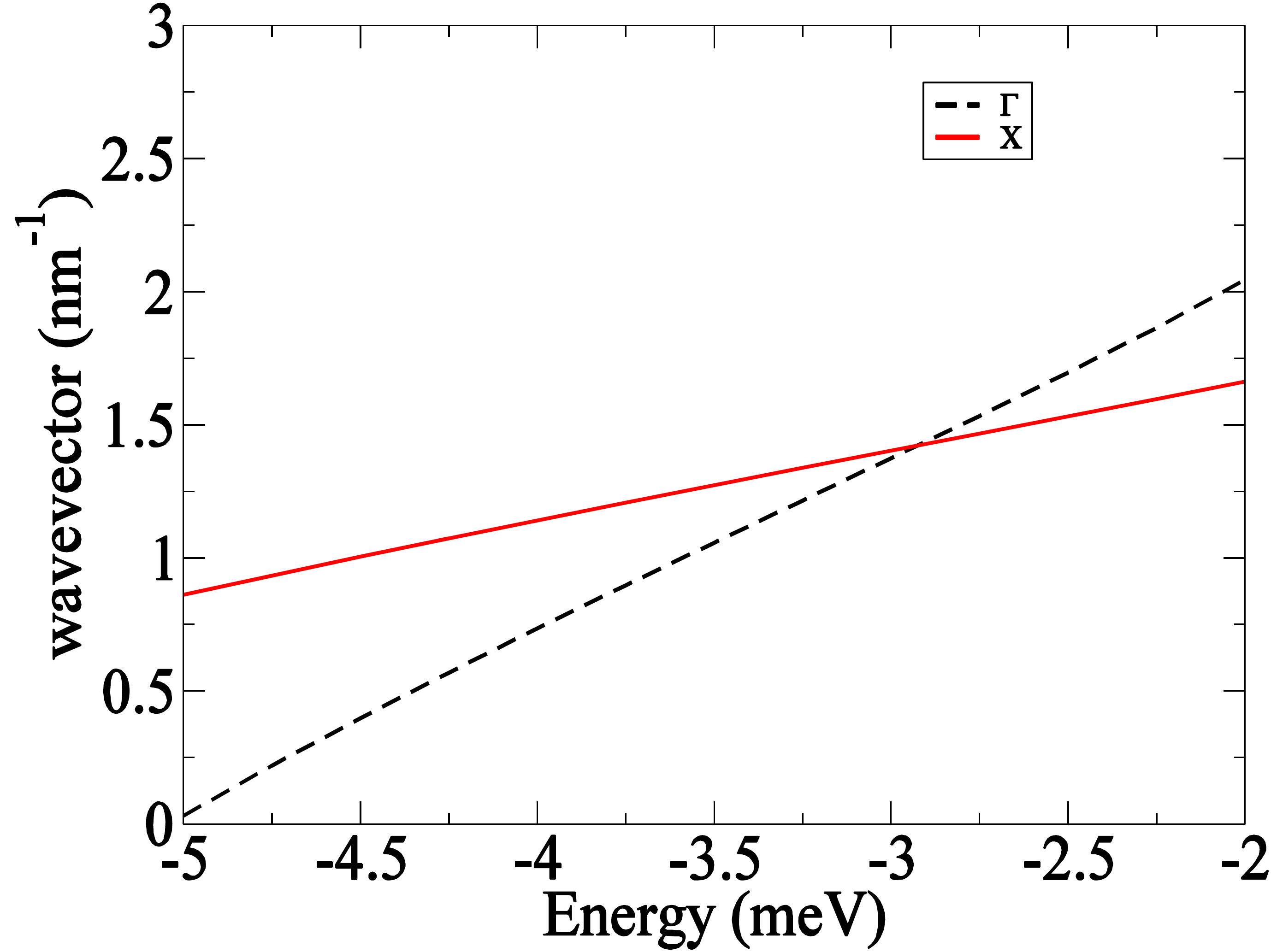}
\caption{(Color online) Fermi wave vector ($k_F$) as a function of Fermi energy ($E_F$) that would be measured by quantum oscillation around each of the Dirac points at $\bar{\Gamma}$ and $\bar{X}$, showing approximately linear dispersions. The average Fermi velocity at the $\bar{\Gamma}$ point is $v^{\bar{\Gamma}}_F=2.3\times 10^3\,m/s$ and that around the $\bar{X}$ point is larger and given by $v^{\bar{X}}_F=5.7 \times 10^3\,m/s$. 
}\label{Fig6}
\end{figure}

Therefore, generically (unless $\eta_d = \eta_f$) there exists an {\it offset} among the position of the Dirac points, residing at the $\bar{\Gamma}$ and $\bar{X}$ points. For the chosen values of the parameters as in Eq. (\ref{smb6para}), all the Dirac points are placed within the bulk insulating gap. However, tuning various parameters in the effective model $H_{bulk}(\bk)$, one can tune various measurable quantities in the bulk such as the hybridization gap, as well as on the surface, such as the energies of the Dirac fermions near different points and the offset among them. In Fig. \ref{Fig5}, we demonstrate the variation of these quantities as a function of a single tuning parameter $\eta_f$, while the rest of the parameters are kept fixed at their values quoted in Eq. (\ref{smb6para}). This plot shows that surface Dirac points can be moved over a certain range in energy and Kondo insulators with different bulk gaps can be realized by changing  band parameters, which may be relevant for other Kondo systems with cubic symmetry.

\begin{figure}[htb]
\includegraphics[width=8.4cm,height=7.0cm]{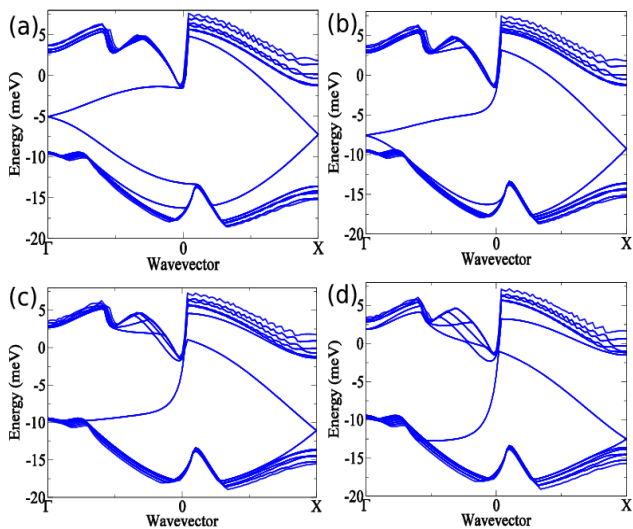}
\caption[] {(Color online) Surface bandstructure for screening length $\lambda_B=2 a$ for various surface potential amplitude (a) $U_0=0$ meV, (b) $U_0=5$ meV, (c) $U_0=10$ meV and (d) $U_0=15$ meV. As the potential increases towards the band-gap, the Dirac point at $\Gamma$ is found to approach the valence band and for sufficiently strong band-bending potential, the Dirac cone at $\Gamma$ disappears into the valence band. The velocity of the $\Gamma$ surface state increases significantly (by an order of magnitude) in (d) relative to (a). The modification to the $X$ point is comparatively minor.}\label{bandbend1}
\end{figure}

Perhaps one of the most intriguing recent experimental results concerns the measurement of the effective mass, and concomitantly the effective Fermi velocity of the surface carriers \cite{qosc}. Quantum oscillations measure the area of the Fermi surface $A(E_F)$ at each of the pockets $\bar{\Gamma}$ and $\bar{X}$ and can be used to estimate the Fermi wave vector 
\beg
k_F(E_F)=\sqrt{A(E_F)/\pi},
\en
where $E_F$ is the Fermi energy. The scaling of $k_F$ near each pockets, as a function of energy of the surface states, obtained from our effective model, are plotted in Fig.~\ref{Fig6}. From this scaling, the Fermi velocity ($v_F$) and corresponding mass ($m$) can be computed since $v_F(E_F)=(\partial k_F/\partial E_F)^{-1}$ and $m(E_F)=k_F(E_F)/v_F(E_F)$. Comparing these results with the quantum oscillations \cite{qosc} and ARPES experiments \cite{exp4, exp5, ARPES2, ARPES3, ARPES4}, we observe that the ratio of the Fermi wave vectors near the $\bar{X}$ point along $k_x$ and $k_y$ directions can be consistent with ARPES measurements and the ratio of the Fermi velocities at the $\bar{X}$ and $\bar{\Gamma}$ points is also consistent with quantum oscillation measurements. In contrast, the typical values of the Fermi velocities and $k_F$, obtained in our calculation are off by more than an order of magnitude than the one extracted from the quantum oscillation and ARPES measurements, respectively.


\section{Band bending}
\label{bending}

\begin{figure}[htb]
\includegraphics[width=8.4cm,height=7.0cm]{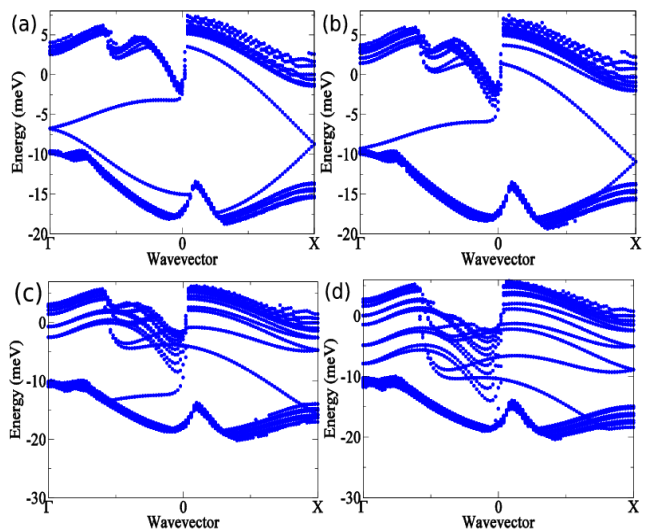}
\caption[] {(Color online) Surface band structure for screening length $\lambda_B=12 a$ for various surface potential amplitude (a) $U_0=2$ meV, (b) $U_0=5$ meV, (c) $U_0=12$ meV and (d) $U_0=20$ meV. In addition to a stronger, but otherwise qualitatively similar, effect on the Dirac cones as in the $\lambda_B=2$ (i.e., Fig.~\ref{bandbend1}) we find the appearance of multiple states in addition to the Dirac cone. These states would likely have small $k_F$.}\label{bandbend2}
\end{figure}

As we have already discussed in the Introduction, our derivation of the effective theory for the surface states is based on the mean-field theory for the interacting Hamiltonian in the bulk. In particular, we have assumed that the hybridization amplitude, as well as the chemical potential, remain spatially homogeneous even close to the surface of the material. An implicit assumption that has been made in this work so far is that the valence of the $f$-ions remains the same both on the surface and in the bulk. Recent experimental studies on SmB$_6$ \cite{RecentCarbon}, however, suggest that the valence state of samarium ion is close to $4f^5$, which is different from the mixed valence state in the bulk. Therefore, coupling between surface and bulk lattice degrees of freedom may play an important role in determining the values of various parameters for the surface electrons \cite{coleman-private, coleman-APS, 1d-bandbending}. We here address this issue by numerically computing the spectrum of the surface states assuming a spatially modulated profile of the chemical potential.

While the tight-binding model captures the topological properties of the surface states, the details of the electronic  structure depend on details of the surface. In particular, generically one can expect a shift in the surface potential from broken bonds at the surface, charged impurities and defects, polar surface termination \cite{damascelli} and surface reconstruction \cite{jeny-hoffman}. We model this surface potential, which requires accounting for self-consistency effects in addition to details of the surface, by an exponential decaying potential with amplitude $U_0$ and decay length scale $\lambda_B$, represented by
\begin{equation}\label{BBPOt}
V(x,y,z)= U_0 \exp[-z/\lambda_B],
\end{equation}
which we add to the tight-binding $H_{bulk}(\bk)$. The strength of the band-bending potential ($U_0$) is not tied with the Kondo insulating gap in the bulk, and therefore it is likely that $U_0 \gg V$.

We first consider the situation of {\it short-ranged} screening by taking $\lambda_B=2 a$ in Fig.~\ref{bandbend1}, where $a$ is the lattice constant. As the potential increases towards the band-gap, the Dirac point at $\bar{\Gamma}$ is found to approach the valence band and for sufficiently strong band-bending potential $(U_0)$, the Dirac cone at $\bar{\Gamma}$ disappears into the valence band. It is interesting to notice that the velocity of the surface states at $\bar{\Gamma}$ point increases significantly as one increases the band-bending potential ($U_0$), in particular by an order of magnitude in $(d)$ relative to $(a)$ in Fig.~7. On the other hand, the modification of the surface states near the $\bar{X}$ point is comparatively minor in comparison to that near the $\bar{\Gamma}$ point.

Next we consider the limit of long-ranged screening by choosing $\lambda_B=12 a$, and the resultant modification in the surface band structure is shown in Fig.~\ref{bandbend2}. A stronger, but otherwise qualitatively similar, effect on the surface Dirac cones is observed in comparison to that for $\lambda_B=2 a$ (i.e., Fig.~\ref{bandbend1}). In addition to the Dirac cones, we also find the appearance of multiple states at the surface when the screening length is large. These states would likely have a small $k_F$. Thus band bending not only significantly renormalizes the Fermi velocity ($v_F$), but also modifies the Fermi wave vector ($k_F$). It is worth pointing out that a realistic strength of the band-bending potential can drag down the Dirac points into the valence band and place it outside the bulk insulating gap, which in SmB$_6$ is $\sim 15$ meV, much smaller than that in Bi$_2$Se$_3$ ($\sim 300$ meV). This may stand as a possible explanation for the absence of surface Dirac points in ARPES measurements \cite{exp4,exp5,ARPES2,ARPES3,ARPES4,ARPES5,ARPES6}.


\section{Discussion and Conclusions}

To conclude, we have derived the effective Hamiltonian for the helical metallic states on the surface of cubic topological Kondo insulators, such as SmB$_6$. The bulk band structure here has been obtained within the mean-field approximation. To derive the surface state Hamiltonian we have projected the inverted even- and odd-parity bands near the high-symmetry points ($X$ points) of the 3D BZ onto the one of the main surfaces. We show that helical Dirac fermionic excitations live around $\bar{\Gamma}$ and $\bar{X}$ points of the surface BZ. While the conical dispersion near the $\bar{\Gamma}$ point is isotropic, that near $\bar{X}$ point is anisotropic. We have also obtained the expressions for the penetration depth and effective Fermi velocities near each of these points. Finally we wish to put forward some connections with recent ARPES \cite{exp4,exp5,ARPES2,ARPES3,ARPES4,ARPES5,ARPES6} and quantum oscillation measurements \cite{oldosi-1, oldosi-2, qosc}.

{\it ARPES}. A number of recent ARPES measurements suggest the existence of an insulating bulk at low temperatures, as well as they have revealed the structure of the surface states in SmB$_6$ \cite{exp4, exp5, ARPES2, ARPES5, ARPES6}. In particular ARPES has shown a circular/isotropic pocket around the $\bar{\Gamma}$ point, and oval/anisotropic pockets in the vicinity of the $\bar{X}$ points of the surface BZ \cite{ARPES2, exp4}. Otherwise, among the energies of the surface Dirac fermions at different points of the BZ, generically there exists an {\it offset}, and that near the $\bar{\Gamma}$ and $\bar{X}$ points are $\sim 18$ meV and $15$ meV, respectively \cite{exp5, ARPES2}. More recent ARPES measurements has also revealed the similar band structure of the surface BZ of SmB$_6$ \cite{exp4, ARPES6}. Fermi surface cuts within the window $\pm 4$ meV, discern pockets near $\bar{\Gamma}$ as well as near the $\bar{X}$ points \cite{exp4}. These observations are in excellent qualitative agreement with our findings, reported in Sec. III. A recent spin-resolved ARPES measurement \cite{spin-ARPES} has confirmed the helical spin-texture for the surface states around the $\bar{X}$ and $\bar{Y}$ points of the surface BZ, we found here.

The helical nature of the surface states can, for example, be established through the mapping of the chirality of the orbital angular momentum using {\it circular dichroism} ARPES measurement \cite{exp5}. Upon mapping the Fermi surfaces using right and left circular polarized light, it has been shown that the ARPES intensities in the portion of the Fermi surface with positive and negative $k_y$ is stronger, respectively. Otherwise, this feature is present near $\bar{\Gamma}$ and $\bar{X}$ points. Consequently, the {\it difference} of the ARPES intensities with right and left circular polarized light clearly discerns an {\it antisymmetric} structure for all the Fermi pockets about the $k_y=0$ axis. Thus circular dichroism ARPES measurements are suggestive of the helical nature of the surface states, which causes locking of spin and orbital angular momenta, yielding {\it helical spin texture} of the surface states of SmB$_6$, shown in Fig. \ref{Fig1}. The helical quasiparticle excitations at low energies near the $\bar{\Gamma}$ and $\bar{X}$ points are, respectively, captured by the low-energy Dirac Hamiltonians $H^{\bar{\Gamma}}_{sur}$ and $H^{\bar{X}}_{sur}$, shown in Eqs. (\ref{gamaHamil}) and (\ref{XHamil}) or (\ref{YHamil}). It is worth mentioning that the circular dichroism ARPES technique has successfully established the helical structure of the surface states in weakly correlated topological insulators, such as Bi$_2$Se$_3$ \cite{circularbi2se3-1, circularbi2se3-2, dicroism-review}.

Recently, an ARPES measurement for another member of the {\it hexaboride family}, YbB$_6$, became available \cite{yb6ARPES, zahedybb6, ybb6-swissARPES}, clearly suggesting the existence of surface states in the vicinity of $\bar{\Gamma}$ and $\bar{X}$ points, similar to SmB$_6$. Furthermore, circular dichroism ARPES measurements with right and left circular polarized light also suggests the helical structure of these surface states, which may arise due to the presence of a topologically nontrivial bulk.  However, it has been argued that YbB$_6$ is possibly not a topological Kondo insulator \cite{zahedybb6}. Nevertheless, our analysis on the band-bending phenomena due to the spatial modulation of the chemical potential may as well be applicable in YbB$_6$, and provide an explanation for the absence of the Dirac points.

{\it Quantum Oscillations}. Recent quantum oscillation measurements also provide valuable insight into the Fermi surface topology of the surface BZ in SmB$_6$. The angular dependence of the {\it out-of-plane component} of {\it magnetoresistance}, measured in the presence of in-plane magnetic fields, discerns a fourfold periodicity, for any field $B>4$T, and at temperature $>5-10$ K \cite{oldosi-1, oldosi-2}, which may arise from the four-fold rotational symmetry among the {\it anisotropic} Fermi pockets around the $\bar{X}$ points in the surface BZ. On the other hand, the isotropic Fermi pocket near the $\bar{\Gamma}$ does not contribute to the oscillation of magnetoresistance.

In addition, quantum oscillation has also been observed in SmB$_6$ using torque magnetometry (de Haas-van Alphen effect) in strong magnetic fields ($B> 5$ T), which through the formation of Landau levels for the two dimensional surface states, yields a very sensitive tool to probe the Fermi surface topology \cite{qosc}. Firstly, the quantum oscillation confirms the existence of two different pockets on $(100)$ surface, which is in accordance with our explicit calculation and also with number of ARPES measurements. The fast Fourier transformation of the torque oscillation gives the oscillation frequencies ($\nu$) for different Fermi pockets, which in turn provides the area of the Fermi pockets (A), since 
\begin{equation}
\nu=\frac{\hbar}{2 \pi e} A,
\end{equation}
where $e$ is electronic charge and consequently the Fermi momentum ($k_F$) \cite{ascroft-mermin}. On the other hand, from the temperature dependence of the oscillation amplitude one finds the effective mass ($m$) of the quasiparticle excitation (Lifshitz-Kosevich formula) \cite{LK}. From the notion of these two quantities, one can find the effective Fermi velocity ($v_F \approx k_F/m$), yielding $\sim (2.9 \pm 0.4) \times 10^5$ m/s near $\bar{\Gamma}$ and $\sim (6.5 \pm 0.21) \times 10^5$ m/s near $\bar{X}$ point \cite{qosc}. The measured values of $v_F$ are roughly two order magnitude larger than their values obtained in ARPES measurement ($0.3$ eV.$\mathring{\mbox{A}}$) \cite{exp4}, which on the other hand, may arise due to the band-bending phenomena \cite{coleman-APS}. Tracking the Landau level index to infinite field limit, which measures the {\it geometric Berry phase}, one obtains an interception $\approx -1/2$ as $H \to \infty$, for both the pockets in residing on $(100)$ plane \cite{qosc}. This observation strongly suggests the existence of topologically protected {\it two component massless Dirac fermionic excitation} around $\bar{\Gamma}$ and $\bar{X}$ points.


\acknowledgements

This work is supported by US-ONR (US) and by LPS-CMTC (B.R. and J. D. S.), DOE-BES DESC0001911 and Simons Foundation (V.G.) and ICAM Senior Fellowship (M.D.). M. D. acknowledges a partial financial support from FCT PTDC/FIS/111348/2009. B. R. is thankful to Pallab Goswami for many useful discussions.


\begin{appendix}

\section{Calculation of the topological invariants}

To compute the topological invariants we need to evaluate the Hamiltonian at the high-symmetry points (HSP) of the BZ. Since the hybridization matrix elements vanish at HSPs, the Hamiltonian can be diagonalized immediately. The resulting band structure consists of four 
(two $d$-like and two $f$-like) doubly degenerate bands: 
\beg\label{eigenvals}
\begin{split}
E^{\pm}_{d}(\bk_m)=&\epsilon^d+\frac{t_d}{2}\left\{(1+\eta_d)(\phi_{1m}+\phi_{2m})\pm\right.\\&\left.(1-\eta_d)\sqrt{(\phi_{1m}-\phi_{2m})^2+4\phi_{3m}^2}\right\}, \\
E^{\pm}_{f}(\bk_m)=&\epsilon^f+\frac{t_f}{2}\left\{(1+\eta_f)(\phi_{1m}+\phi_{2m})\pm\right.\\&\left.(1-\eta_f)\sqrt{(\phi_{1m}-\phi_{2m})^2+4\phi_{3m}^2}\right\},
\end{split}
\en
where
\beg
\phi_{\alpha m}=\phi_\alpha(\bk_m), \quad \alpha=1,2,3.
\en
In the basis shown in Eq.~(\ref{basis}) the inversion operator is
\beg
\hat{P}=\hat{\sigma}_z\otimes\hat{\tau}_0\equiv\textrm{diag.}(1,1,-1,-1).
\en 
Consider the Hamiltonian $\hat{H}_{bulk}(\bk_m)=\hat{H}_m$ from Eq. (\ref{Hbulk}) evaluated in the HSP in the basis of the eigenstates corresponding to the eigenvalues (\ref{eigenvals}):
\beg\label{Hm}
\hat{H}_m=\textrm{diag}(E^+_{d}(\bk_m), E^-_{d}(\bk_m), E^+_{f}(\bk_m), E^-_{f}(\bk_m)).
\en
It follows that Eq. (\ref{Hm}) can be written as a sum of four operators:
\begin{eqnarray}
\hat{H}_m= \nonumber \\
\frac{\hat{\sigma}_z\otimes\hat{\tau}_z}{4}\left[E^+_{d}(\bk_m)-E^-_{d}(\bk_m)-E^+_{f}(\bk_m)+E^-_{f}(\bk_m)\right] \nonumber \\
+ \frac{\hat{\sigma}_0\otimes\hat{\tau}_z}{4}\left[E^+_{d}(\bk_m)-E^-_{d}(\bk_m)+E^+_{f}(\bk_m)-E^-_{f}(\bk_m)\right] \nonumber \\
+\frac{\hat{\sigma}_0\otimes\hat{\tau}_0}{4}\left[E^{+}_{d}(\bk_m)+E^{-}_{d}(\bk_m)+E^{+}_{f}(\bk_m)+E^{-}_{f}(\bk_m)\right] \nonumber \\
+\frac{\hat{\sigma}_z\otimes\hat{\tau}_0}{4}\left[E^+_{d}(\bk_m)+E^-_{d}(\bk_m)-E^+_{f}(\bk_m)-E^-_{f}(\bk_m)\right]. \nonumber \\
\end{eqnarray}
Note that the last term in this expression is proportional to the parity operator. 

To compute the invariant we need to consider the bands which are occupied at least at one point of the BZ. Since the $d$-band $E^+_{d}(\bk_m)$ is highest energy it remains unoccupied at all points of the BZ and therefore it can be ignored. For the remaining three bands we
can set the parity eigenvalues to
\beg
\begin{split}
\delta_m=+1: \quad E^-_{d}(\bk_m)>E^-_{f}(\bk_m)>E^+_{f}(\bk_m), \\
\delta_m=-1: \quad E^-_{f}(\bk_m)>E^-_{d}(\bk_m)>E^+_{f}(\bk_m), \\
\delta_m=-1: \quad E^-_{f}(\bk_m)>E^+_{f}(\bk_m)>E^-_{d}(\bk_m).
\end{split}
\en
Note that $E^-_{f}>E^+_{f}$ since we are considering $d$-electron bands and $f$-hole bands to ensure that insulating gap does not vanish
anywhere in the BZ. Therefore, the parity eigenvalue is
\beg
\delta_m=\textrm{sign}\left[E^-_{d}(\bk_m)-E^-_{f}(\bk_m)\right].
\en
Then, the topological invariant is determined by 
\beg
(-1)^\nu=\prod\limits_{m=1}^8\delta_m.
\en
The dependence of $\delta_m$ and $\nu$ on the microscopic parameters such as bare hybridization $V$ and $f$-level energy $\varepsilon_f$ has been analyzed in Refs. \onlinecite{Takimoto}, \onlinecite{CTKI}. It was found that strong topological Kondo insulator, $\nu=-1$, is realized for a wide range of values of $V, \varepsilon_f$. 

\end{appendix}


\end{document}